\documentclass{ws}

\def\etal {{\it et al.}}
\def\lvgap{Ref.\ \refcite{lvgap}}
\def\rlvgap{Reference \refcite{lvgap}}

\def\et{\eta}

\def\mn{{\mu\nu}}

\def\frac#1#2{{\textstyle{{#1}\over {#2}}}}

\def\lsim{\mathrel{\rlap{\lower4pt\hbox{\hskip1pt$\sim$}}
    \raise1pt\hbox{$<$}}}
\def\gsim{\mathrel{\rlap{\lower4pt\hbox{\hskip1pt$\sim$}}
    \raise1pt\hbox{$>$}}}
\def\sqr#1#2{{\vcenter{\vbox{\hrule height.#2pt
         \hbox{\vrule width.#2pt height#1pt \kern#1pt
         \vrule width.#2pt}
         \hrule height.#2pt}}}}

\def\etal{{\it et al.}}

\def\pt#1{\phantom{#1}}

\def\ab{\overline{a}{}}

\def\cb{\overline{c}{}}

\def\sb{\overline{s}{}}

\def\twiddle{\lower4pt\hbox{\hskip-0pt{$\widetilde{}$}}}
\def\m@th{\mathsurround=0pt}
\def\cmapstochar{\mathrel{\rlap{
  \lower0.1pt\hbox{\hskip-1.75pt{$\mapstochar$}}}
  \raise0pt\hbox{\hskip2.5pt{$\twiddle$}}}}
\def\notsimfill{$\m@th\cmapstochar$}
\def\scroodle#1{\vbox{\ialign{##\crcr\notsimfill\crcr
  \noalign{\kern-4pt\nointerlineskip}
   $\hfil\displaystyle{#1}\hfil$\crcr}}}

\def\cmapstocharbig{\mathrel{\rlap{
  \lower0.1pt\hbox{\hskip0.25pt{$\mapstochar$}}}
  \raise0pt\hbox{\hskip4.5pt{$\twiddle$}}}}
\def\notsimfillbig{$\m@th\cmapstocharbig$}
\def\scroodlebig#1{\vbox{\ialign{##\crcr\notsimfillbig\crcr
  \noalign{\kern-4pt\nointerlineskip}
   $\hfil\displaystyle{#1}\hfil$\crcr}}}

\def\af{(a_{\rm{eff}})}

\def\afb{(\ab_{\rm{eff}})}

\def\lrpartial{\raise 1pt\hbox{$\stackrel\leftrightarrow\partial$}}

\def\summarya{XIV}
\def\summaryb{XV}

\newcommand{\beq}{\begin{equation}}
\newcommand{\eeq}{\end{equation}}
\newcommand{\bea}{\begin{eqnarray}}
\newcommand{\eea}{\end{eqnarray}}
\newcommand{\bit}{\begin{itemize}}
\newcommand{\eit}{\end{itemize}}

\def\pno#1{PNO(#1)}

\begin{document}

\title{LORENTZ SYMMETRY AND MATTER-GRAVITY COUPLINGS}

\author{JAY D.\ TASSON}

\address{Department of Physics, Whitman College\\
Walla Walla, WA 99362, USA\\
E-mail: tassonjd@whitman.edu}

\begin{abstract}
This proceedings contribution summarizes recent investigations
of Lorentz violation in matter-gravity couplings.
\end{abstract}

\bodymatter

\section{Introduction}

In spite of the many high-sensitivity investigations 
of Lorentz violation\cite{tables}
performed in the context of 
the fermion sector of the minimal Standard-Model Extension (SME)
in Minkowski spacetime,\cite{ck}
only about half of the coefficients for Lorentz violation
in that sector have been investigated experimentally.
\rlvgap\ establishes a methodology
for obtaining sensitivities to some of these open parameters
by considering gravitational couplings
in the fermions sector of the SME,\cite{akgrav}
extending pure gravity work.\cite{pureg}
Of particular interest
are the $\ab_\mu$ coefficients
for baryons and charged leptons,
which are unobservable in principle
in Minkowski spacetime,
but could be relatively large 
due to gravitational countershading.\cite{akjt}

The first half of \lvgap\ develops the necessary theoretical results
for the analysis of Lorentz violation in matter-gravity couplings.
Those results are summarized in Sec.\ \ref{theory}
below,
while Sec.\ \ref{expt} summarizes 
the experimental predictions provided in the second half of that work.

\section{Theory}
\label{theory}

The theoretical portion of \lvgap\
addresses a number of useful conceptual points 
prior to developing the necessary results
for experimental analysis.
This includes a discussion 
of the circumstances under which relevant types of Lorentz violation
are observable in principle.
It turns out that the $\ab_\mu$ coefficient,
which can be removed from the single fermion theory
in Minkowski spacetime
via a spinor redefinition
cannot typically be removed in the presence of gravity.\cite{akgrav}
This makes it an interesting case for study
in the remainder of \lvgap.
A coordinate choice
that can be used to fix the sector of the theory
that defines isotropy is also discussed
and ultimately used to take the photon sector
to have $\et_\mn$ as the background metric.

Another issue is the development
of general perturbative techniques to treat
the fluctuations in the coefficient fields
in the context of matter-gravity couplings.
Two notions of perturbative order are introduced.
One, denoted O($m,n$),
tracks the orders in Lorentz violation and in gravity,
where the first entry represents the order 
in the coefficients for Lorentz violation
and the second represents the order in the metric fluctuation $h_\mn$.
The secondary notion of perturbative order,
denoted \pno{$p$},
tracks the post-newtonian order. 
The goal of \lvgap\ is to investigate
dominant Lorentz-violating implications
in matter-gravity couplings,
which are at O(1,1).

\rlvgap\ provides the necessary results
to analyze experiments at a variety of levels
while working toward the classical nonrelativistic equations of motion,
which are most relevant for many of the experiments 
to be considered.
Development of the quantum theory of the gravity-matter system
provides the first step.
Starting from the field-theoretic action,
the relativistic quantum mechanics 
in the presence of gravitational fluctuations
and Lorentz violation is established
after investigating methods of identifying
an appropriate hamiltonian
in the presence of an effective inverse vierbein $E_\mu^{\pt{\mu} 0}$.
The explicit form of the relativistic hamiltonian
involving all coefficients for Lorentz violation
in the minimal QED extension is provided. 
Attention is subsequently specialized to the study
of spin-independent Lorentz-violating effects,
which are governed by the coefficient fields $\af_\mu$, $c_\mn$
and the metric fluctuation $h_\mn$.
Analysis then proceeds to the 
nonrelativistic quantum hamiltonian 
via the standard Foldy-Wouthuysen procedure.

While the quantum mechanics above
is useful for analysis of quantum experiments,
most measurements of gravity-matter couplings
are performed at the classical level.
Thus the classical theory\cite{nr} 
associated with the quantum-mechanical dynamics
involving nonzero $\af_\mu$, $c_\mn$, and $h_\mn$
is provided at leading order in Lorentz violation 
both for the case of 
the fundamental particles appearing in QED
and for bodies involving many such particles.
These results enable the derivation
of the modified Einstein equation
and the equation
for the trajectory of a classical test particle.
Solving for the trajectory
requires knowledge of the coefficient and metric fluctuations.
A systematic methodology 
for calculating this information
is provided,
and general expressions 
for the coefficient and metric fluctuations
to O(1,1)
in terms of various gravitational potentials
and the background coefficient values $\afb_\mu$ and $\cb_\mn$
are obtained.
Bumblebee models are considered
as an illustration
of the general results.

\section{Experiments}
\label{expt}

A major class of experiments
that can achieve sensitivity to coefficients $\afb_\mu$ and $\cb_\mn$
involve laboratory tests with ordinary neutral matter.
Tests of this type are analyzed via 
the \pno3 lagrangian describing the dynamics of a test body
moving near the surface of the Earth
in the presence of Lorentz violation.
The analysis reveals that the gravitational force acquires tiny corrections
both along and perpendicular to the usual free-fall trajectory
near the surface of the Earth,
and the effective inertial mass of a test body
becomes a direction-dependent quantity.
Numerous laboratory experiments
sensitive to these effects 
are considered.
The tests can be classified as either 
gravimeter or Weak Equivalence Principle (WEP) experiments
and as either force-comparison or free-fall experiments
for a total of 4 classes. 

Free-fall gravimeter tests
monitor the acceleration of freely falling objects
and search for the characteristic time dependence 
associated with Lorentz violation.
Falling corner cubes\cite{fc} and matter interferometry\cite{ai,aif}
provide examples of such experiments
and are discussed in \lvgap.
Force-comparison gravimeter tests
using equipment such as superconducting gravimeters
are also studied.\cite{sg}
Note that the distinction, force comparison verses free fall,
is important due to the potential Lorentz-violating misalignment
of force and acceleration.
Making direct use of the flavor dependence 
associated with Lorentz-violating effects
implies signals in WEP tests.
A variety of free-fall WEP tests
are considered including those using
falling corner cubes,\cite{fcwep}
atom interferometers,\cite{aiwep,aif}
tossed masses,\cite{tossed}
balloon drops,\cite{balloon}
drop towers,\cite{bremen}
and sounding rockets,\cite{srpoem}
along with force-comparison WEP tests
with a torsion pendulum.\cite{tpwep}
For all of the tests considered,
the possible signals for Lorentz violation
are decomposed according to their time dependence,
and estimates of the attainable sensitivities are obtained.

Satellite-based WEP tests,\cite{space}
which offer interesting prospects for improved sensitivities
to Lorentz violation,
are also discussed in detail.
The signal is decomposed by frequency
and estimated sensitivities 
are obtained.

The experimental implications of Lorentz violation 
in the gravitational couplings 
of charged particles, antimatter,
and second- and third-generation particles
are also studied.
These tests are experimentally challenging,
but can yield sensitivities to Lorentz and CPT violation 
that are otherwise difficult or impossible to achieve.
Possibilities
including charged-particle interferometry,\cite{chargeai}
ballistic tests with charged particles,\cite{charge}
gravitational experiments with antihydrogen,\cite{anti}
and signals in muonium free fall\cite{muon}
are discussed.
Simple toy models are used to illustrate 
some features of 
antihydrogen tests.

Solar-system tests of gravity
including lunar and satellite laser ranging tests\cite{llr}
and measurements of the precession of the perihelion
of orbiting bodies\cite{peri} are also considered.
The established advance of the perihelion
for Mercury and for the Earth is used to obtain constraints
on combinations of $\afb_\mu$, $\cb_\mn$, and $\sb_\mn$,
which provides the best current sensitivity to $\afb_J$.

A final class of tests 
involves the interaction of photons with gravity.
Signals arising 
in measurements of the time delay,
gravitational Doppler shift,
and gravitational redshift,
are considered along with
comparisons of the behaviors of photons and massive bodies.
Implications for a variety of existing and proposed experiments
and space missions are considered.\cite{photon}
 
Existing and expected sensitivities from the experiments and observations
summarized above are collected in
Tables \summarya\ and \summaryb\
of \lvgap.
These sensitivities
reveal excellent prospects for using matter-gravity couplings 
to seek Lorentz violation.
The opportunities for measuring 
the countershaded coefficients $\afb_\mu$
are particularly interesting
in light of the fact that
these coefficients typically cannot be detected 
in nongravitational searches.
Thus the tests proposed in \lvgap\
offer promising new opportunities
to search for signals of new physics,
potentially of Planck-scale origin.

\end{document}